# Zwitterionic Polymer Coatings with Compositional Gradient for Stable and Substrate-Independent Biofouling Deterrence via All-Dry Synthesis

Pengyu Chen[a], Harry Shu[a], Wenjing Tang[a], Christina Yu[a], Rong Yang[a*]

[a.] *Robert Frederick Smith School of Chemical and Biomolecular Engineering, Cornell University, Ithaca, NY, 14853, USA*

*\* Corresponding author: ryang@cornell.edu*

## Abstract

Biofouling represents a critical challenge in marine transportation, healthcare, and food manufacturing, among other industries, as it promotes contamination and increases maintenance costs. Zwitterionic polymers, known for their exceptional antifouling properties, offer a promising solution for biofouling deterrence. Despite the rapid development of zwitterionic polymers in recent years, the design rules, especially concerning the choice of cationic moieties to optimize biofouling deterrence, remain elusive. In this study, we leveraged a versatile all-dry synthesis scheme to achieve a selection of 9 zwitterionic polymers, 5 of which are unprecedented for this synthesis paradigm, thus systematically unraveling that molecular design rule. Notably, we developed a synthesis strategy to enable nanoscale compositional gradient along the coating cross-section, which ensures the robustness of the zwitterionic polymer coatings irrespective of the choice of cation-anion combinations. That robustness is enabled by an organosilicon-based layer at the coating-substrate interface, which simultaneously enhances coating adhesion and chemical stability while ensuring high concentration of zwitterionic moieties at the polymer-liquid interface to maximize biofouling deterrence. The antifouling efficacy was assessed using biofilms of *Pseudomonas aeruginosa* or *Bacillus subtilis*. All coatings demonstrated antifouling efficacy, with a novel zwitterionic polymer comprising a combination of imidazolium and carboxyl groups achieving the greatest antibiofilm effects, which we attributed to the strong hydration. This study highlights the coating architecture, i.e., one with nanoscale gradient and varying crosslinking densities, as a valid strategy to render zwitterionic polymers robust coatings and the imidazolium-based carboxybetaine as a promising next-generation antibiofouling chemistry.

**Introduction**

Biofouling, especially that caused by persistent microbial biofilms, pose a significant challenge across various sectors, from marine transportation(*1*) to healthcare(*2*), and to food manufacturing(*3*), due to their roles in spreading infections, contaminating surfaces, and increasing maintenance costs. In healthcare settings, biofilms are one of the major contributors to hospital-acquired infections, costing over two billion dollars of additional medical expense and over thousands of deaths annually(*4*). Beyond healthcare, biofilms in food manufacturing facilities result in life-threatening contaminations and enormous waste(*5*). These examples underscore the need for effective biofilm control strategies.

Zwitterionic polymers, with their highly hydrophilic and charge-neutral properties, have emerged as a promising antifouling material(*6*). By forming a strong hydration layer on the surface, zwitterionic polymers effectively repel nonspecific binding of foulers across the length scales of nanometer to millimeter(*7, 8*). Despite their rapid development and deployment over the past decade, a small number of zwitterionic chemistries have been developed, leading to limited understanding of their molecular design rules in the context of biofouling deterrence. To unravel how the choice of cationic group in a zwitterionic moiety impacts the antifouling performance, we leveraged a versatile all-dry synthesis scheme to achieve a selection of 9 zwitterionic polymers, 5 of which are unprecedented for this synthesis paradigm(*9-12*).

Our synthesis approach combines initiated Chemical Vapor Deposition (iCVD) with selective derivatization, enabling zwitterionic polymer coatings with nanoscale compositional gradient along the cross-section to simultaneously enhance durability and antifouling efficacy(*13*). The gradient coating composition transitions from an organosilicon-based crosslinker at the coating-substrate interface, which provides strong adhesion and durability, to a zwitterionic topmost interface that confers antifouling properties. The use of iCVD enables precise application of ultrathin films in a substrate-independent and conformal fashion, pointing to a versatile and scalable antifouling solution for a wide range of applications(*14*).

Among the zwitterionic polymers synthesized and tested, an unprecedented zwitterionic coating, i.e., gradient imidazolium-based carboxybetaine (VI-PL) coating, demonstrated the greatest antibiofilm efficacy against *Pseudomonas aeruginosa* (a Gram-negative strain) and *Bacillus subtilis* (a Gram-positive strain). Compared to established zwitterionic antifouling coatings, such as the pyridinium-based sulfobetaine(*15*), VI-PL further reduced biofilm formation by over 90%, which we attributed to its strong hydration and distinct molecular charge structure. This study underscores the effectiveness of zwitterionic polymer coatings with compositional gradient as durable, high-performance antifouling materials, and the importance of molecular understanding in the design of next-generation antifouling polymers.

**Materials and Methods**

Material synthesis. Initiator (tert-butyl peroxide (TBPO, Sigma-Aldrich, 98%)), monomers (1,3,5,7-tetramethyl-1,3,5,7-tetravinyl cyclotetrasiloxane (V4D4, Ambeed, 97%), 4-vinyl pyridine (4VP, Sigma-Aldrich, 95%), 1-vinyl imidazole (1VI, Sigma-Aldrich, 99%), and 2-(dimethylamino)ethyl acrylate (DMAEA, Sigma-Aldrich, 98%)), were used without further purification.

Zwitterionic polymer coatings with compositional gradient was synthesized using a two-step synthesis scheme on Si wafers (P/Boron<100>, Purewafer) and 96 well plates. In the first step, iCVD was performed in a custom-built vacuum reactor (Kurt Lesker). During iCVD, 1.0 sccm TBPO was introduced to the reactor at room temperature through a mass flow controller for all the depositions. Thermal activation of the initiator was provided by resistively heating a 0.5 mm nickel/chromium filament (80% Ni/ 20% Cr, Goodfellow) mounted as a parallel filament array. Filament heating was controlled by a DC power supply to maintain a temperature of 250°C throughout the deposition. The deposition stage that was kept at 35°C using a recirculating chiller. The vertical distance between the filament array and the stage was 2 cm. Monomers were heated to desired temperatures (75°C for V4D4, 55°C for 4VP, 65°C for 1VI, 60°C for DMAEA) in their respective glass jars to create sufficient pressure to drive the vapor flow. To create the gradient, a 100-nm poly(V4D4) was deposited using a V4D4 flow rate of 0.2 sccm and a chamber pressure of 200 mTorr. Subsequently, the flow rate of V4D4 was gradually decreased to 0 and that of 4VP or 1VI or DMAEA was gradually increased to desired values of 4 sccm, 2 sccm, and 3 sccm, respectively. The total chamber pressure was gradually increased to 500 mTorr. In situ interferometry with a HeNe laser source (wavelength = 633 nm, JDS Uniphase) was used to monitor the film growth on a Si substrate to be~ 200 nm. A more accurate film thickness was measured post-deposition using a J. A. Woollam Alpha-SE spectroscopic ellipsometry at three different incidence angles (65°, 70°, 75°).

Materials characterization. Fourier transform infrared (FTIR) measurements were performed on a Bruker Vertex V80v vacuum FTIR system in transmission mode. A deuterated triglycine sulfate (DTGS) KBr detector was used over the range of 600−4000 cm$^{-1}$ with a resolution of 4 cm$^{-1}$. Each spectrum was averaged over 64 scans to obtain a sufficient signal-to-noise ratio. All the spectra were processed by background subtraction (subtracting the spectrum of a bare Si wafer) and baseline-correction.

In XPS, samples were analyzed using a Surface Science Instruments SSX-100 ESCA Spectrometer with operating pressure ca. 1x10$^{-9}$ Torr. Monochromatic Al Kα x rays (1486.6 eV) were used to generate photoelectrons over an 800-μm-diameter area on the surface of the samples. Photoelectrons were collected at a 55° emission angle with a source-to-analyzer angle of 70°. A hemispherical analyzer determined electron kinetic energy using a pass energy of 150 eV for wide/survey scans. All the samples were stored under vacuum at room temperature for a week before XPS analysis.

Surface roughness and morphology were measured using a Cypher S AFM. Scans were recorded across a 10 × 10 μm region at a frequency of 0.5 Hz and in tapping mode.

Water contact angle measurements were performed using a Rame-Hart Model 500 goniometer equipped with an automated liquid dispenser. Static water contact angle measurements were performed using 2 μL droplets, dispensed on uncoated or coated silicon wafers.

<u>Bacteria culture and assays to assess surface adhesion, biofilm formation, and metabolic activities.</u> Frozen cultures of *P. aeruginosa*, strain PAO1, and *B. subtilis*, strain DS2569, were stored in LB with 20% (v/v) glycerol at −80°C. A swab of the frozen culture was recovered on LB agar at 37°C for 24 h. Single colonies were inoculated into LB broth and cultured at 37°C for 16 h.

Biofilm formation was assessed in 96-well plates. The coated and uncoated samples were rinsed with DI water, sterilized by UV radiation for 30 minutes, followed by compressed air blow-drying before use. During the assays, the aforementioned 16-h culture was diluted to a concentration of ~$10^7$ CFU/mL (i.e., $OD_{600}$=0.2). The entire 96-well microplates were incubated (37°C) for 24 hours to ensure biofilms fully developed. Biofilms that attached to the microplates were rinsed 3-4 times with DI water to remove the loosely attached bacteria, and stained using 0.1% crystal violet for 10 min. The crystal violet solution was removed by rinsing each well 3-5 times until the liquid in each well became clear and free of color. The microplate was dried in the air at room temperature for 24 h to remove residual water and biofilm formation was subsequently quantified by extracting the absorbed crystal violet from the biofilm using acetic acid (30 vol%) and characterized spectrophotometrically by measuring the $OD_{570}$ using a microplate reader (Infinite M1000Pro, Tecan).

**Results and Discussion**

The gradient zwitterionic polymer coatings were synthesized using a two-step procedure (**Figure 1 (a)**). In the first step, initiated Chemical Vapor Deposition (iCVD) polymerization was performed to form a conformal polymer coating with compositional gradients in the coating cross-section. The base layer comprises poly(1,3,5,7-tetramethyl-1,3,5,7-tetravinyl cyclotetrasiloxane) (pV4D4), which provides strong adhesion, high cross-linking density (for durability), and unreacted vinyl groups that serve as covalent anchoring points for the zwitterionic top layer. The composition of the vapor-phase precursors was tuned continuously (via mass flow controllers) during the iCVD polymerization, from V4D4 to 4-vinyl pyridine

(4VP)/ 1-vinyl imidazole (1VI)/ 2-(dimethylamino)ethyl acrylate (DMAEA), such that the coating composition varies from pV4D4 to a nitrogen-containing top layer (**Figure 1 (b)**). In the second step, the N-containing moieties in the top layer were derivatized to form zwitterionic moieties through an $S_N2$ nucleophilic substitution(*16*). This reaction was performed with vaporized derivatizing agents, including 1,4-butane sultone (BS), 1,3-propane sultone (PS), or propiolactone (PL) (**Figure 1 (c)**). These choices of the derivatizing agent enabled unprecedented zwitterionic polymer structures to be synthesized in an all-dry fashion, enabling carboxybetaines and sulfobetaines with tunable cation-anion coupling.

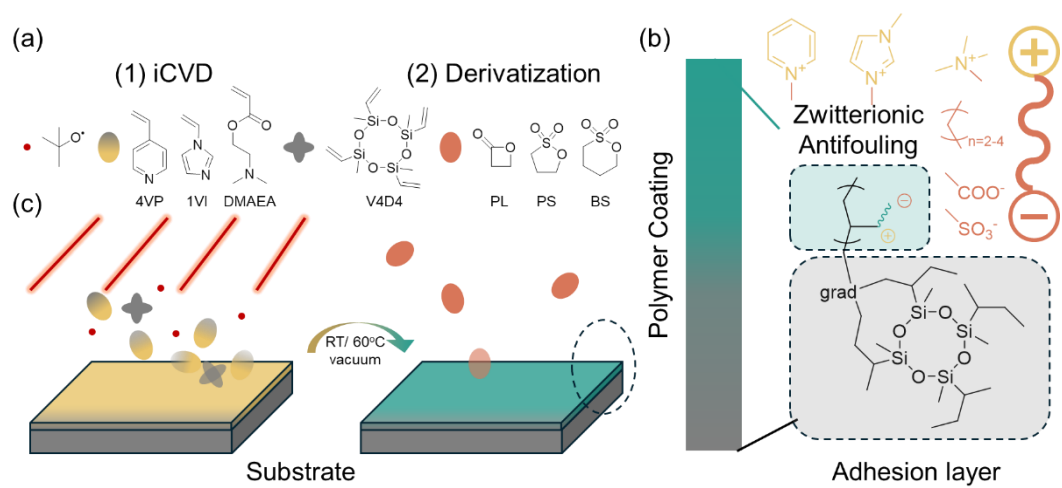

**Figure 1. Schematic representation of the surface design and synthesis of a library of zwitterionic polymer coatings with compositional gradient.** (a) Precursors used in the synthesis, including monomers for iCVD (1,3,5,7-tetramethyl-1,3,5,7-tetravinyl cyclotetrasiloxane (V4D4), 4-vinyl pyridine (4VP), 1-vinyl imidazole (1VI), 2-(dimethylamino)ethyl acrylate (DMAEA)) and derivatization agents (1,4-butane sultone (BS), 1,3-propane sultone (PS), and propiolactone (PL)). (b) Illustration of the compositional gradient of the polymer coatings, featuring an adhesion-promoting pV4D4 base layer and an antifouling zwitterionic top layer. (c) Schematic of the two-step synthesis approach: iCVD enables the compositional gradient and an all-dry derivatization creates zwitterionic moieties with 9 combinations of cations and anions.

The chemical composition and film architecture of the gradient polymer coatings (denoted as V4D4-*grad*-zwitterion precursor-derivatizing agent) were verified using Fourier transform infrared spectroscopy (FTIR) and X-ray photoelectron spectroscopy (XPS). The polymer coatings exhibited characteristic FTIR peaks, confirming the presence of key components (**Figure 2(a), (b), and (c)**). The strong absorption band at 1100 cm$^{-1}$, corresponding to the Si-O stretching vibration, validated the inclusion of V4D4 in the polymer structure(*17*). The V4D4 repeat units also led to unreacted vinyl groups at 2870 cm$^{-1}$(*18*). Upon derivatization with BS or PS, a distinct absorption peak at 1037 cm$^{-1}$ was observed, indicative of SO$_3^-$ groups and confirming the successful transformation of tertiary amines into zwitterionic moieties(*15*). Similarly, derivatization with propiolactone (PL) led to a peak at 1700 cm$^{-1}$, attributed to the C=O stretching vibration(*19*).

XPS analysis corroborated the FTIR findings and provided quantitative insights into the gradient structure and conversion rate of N-containing groups within the zwitterionic polymer coatings. Depth profiling by XPS revealed a progressive increase in the Si atomic ratio, alongside a corresponding decrease in the N atomic ratio. We converted that N atomic ratio into a fraction of the N-containing monomer in the gradient coating (**Figure 2(d)**), confirming the formation of a compositional gradient on the nanoscale along the coating cross-section. Surface composition of the N-containing groups was notably high, with the molar fraction reaching 95% for 4VP, 94% for 1VI, and 91% for DMAEA in the topmost layer. High-resolution N1s scans further confirmed the obtainment of zwitterionic moieties, as evidenced by the distinct peak at 401.5 eV, representing positively charged nitrogen (N$^+$) (**Figure 2(e)** and **Supplementary Figure S2**). (The peak at 398–400 eV represents the precursor N(*15*).) The surface composition of the zwitterionic moieties varied between 44% and 85% across different polymer coatings, which we attributed to factors such as steric hindrance, nucleophilicity, and the volatility of the derivatization agents that affected the conversion rate of the derivatization reaction. For instance, BS has lower volatility and smaller ring strain, and thus lower reactivity and a smaller conversion rate compared to PS(*20*); PL undergoes spontaneous ring-opening polymerization, which had to be mitigated by reducing the temperature and duration of the derivatization reaction (to room temperature and 30 minutes), resulting in a low conversion rate(*21*). Despite the variability, the surface composition of the zwitterionic moieties reported here is significantly higher than those previously reported organosilicon-based polymer coating, i.e., < 40% (*8*)(**Figure 2(f)**). Moreover, similar surface compositions were obtained for zwitterionic moieties bearing the same anionic group, which allowed the comparison of the antifouling performance upon varying the cationic group.

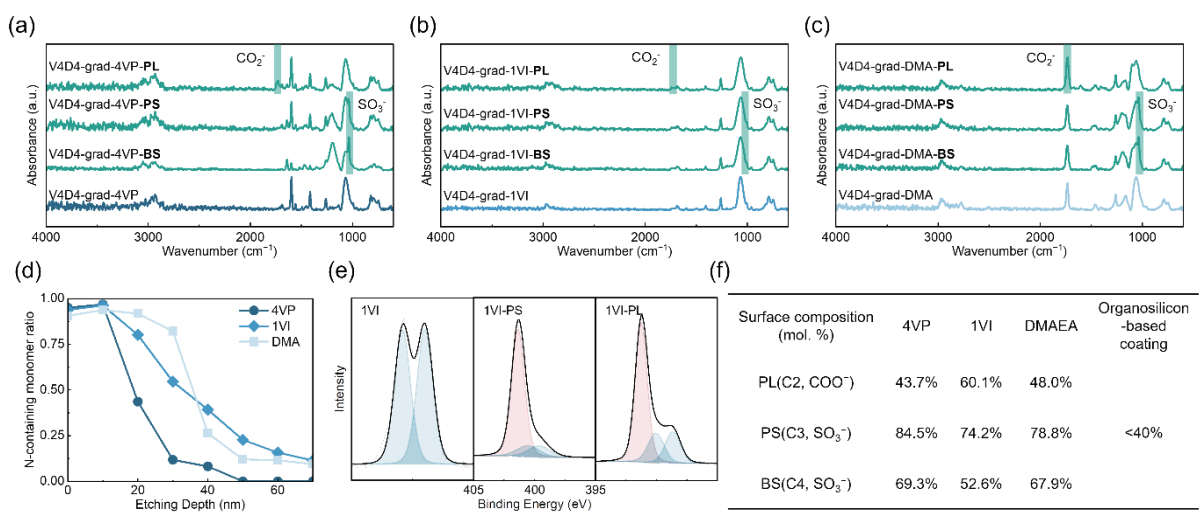

**Figure 2. Chemical characterizations of the zwitterionic polymer coatings with nanoscale compositional gradient.** FTIR spectra of the gradient coatings that are (a) 4VP-based, (b) 1VI-based, and (c) DMAEA-based polymers before and after the derivatization by PL, BS, and PS. (d) XPS depth profiling analysis of a 200-nm-thick gradient polymer coating, showing a compositional transition from a pV4D4 base layer to a N-containing top layer. (e) High-resolution N1s scans on the 1VI-based polymers that are un-derivatized (left) and derivatized with PS (middle) and PL (right). (f) The compositions of the zwitterionic moieties in the topmost layer of the gradient coatings.

Atomic force microscopy (AFM) was used to analyze the surface morphology of the gradient zwitterionic coatings (prepared on Si wafers). As shown in **Figure 3 (a) and (b)**, most polymer coatings exhibited a root-mean-square (RMS) roughness below 1.5 nm, indicating a smooth and uniform surface morphology, which is essential for maintaining the antifouling properties and ensuring consistency in performance. The only exception is the DMAEA-based carboxybetaine, displaying an RMS roughness of 6.8 ± 0.2 nm. The smooth surface morphology of the coatings confirms that the all-dry synthesis effectively preserved the structure of the underlying substrate, and the compositional gradient did not lead to phase separation or other morphological features.

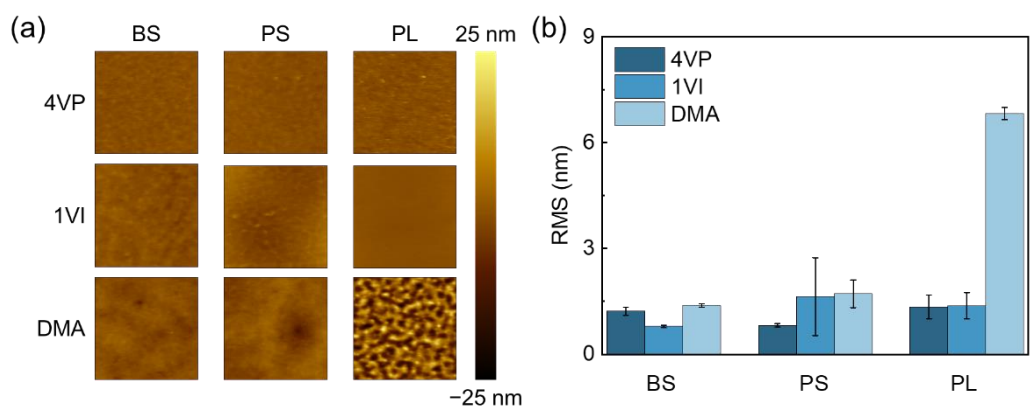

**Figure 3.** (a) AFM images of the zwitterionic coatings (on Si wafer) with compositional gradients, illustrating the smooth surface morphology. (b) The root mean square (RMS) roughness calculated from the AFM images. N = 4, error bar = standard deviation.

Water contact angle (WCA) measurements were conducted to evaluate the wettability and surface energy of the gradient coatings (**Figure 4(a)** and Supplementary **Figure S3**). The WCA of the underivatized gradient coatings ranged between 45° and 75°. The mild hydrophilicity was attributed to the presence of polar N-containing groups. The WCA decreased after the derivatization, indicating enhanced hydration upon converting the N-containing groups to zwitterionic ones. As shown in **Figure 4(b)**, for the 4VP-based coatings, the WCA decreased from 48 ± 1° to 19 ± 1° (4VP-BS), 12 ± 2° (4VP-PS), and 21 ± 1° (4VP-PL); the 1VI-based coating exhibited a WCA decrease from 76 ± 1° to 63 ± 1° (1VI-BS), 47 ± 1° (1VI-PS), and 27 ± 0.3° (1VI-PL); the DMAEA-based coating exhibited WCA decreases from 65 ± 5° to 32 ± 0.3° (DMA-BS), 21 ± 2° (DMA-PS), and 14 ± 0.5° (DMA-PL). By comparing the derivatization products of PS and BS, we concluded that reducing the carbon spacer length enhances hydrophilicity; whereas the comparison between PS and PL indicated that strong anions (i.e., COO⁻) increases hydrophilicity. While the 4VP-based zwitterions represent an outlier to the latter, the general trends observed here align with previous studies indicating that carboxybetaine exhibits stronger hydration than sulfobetaine.

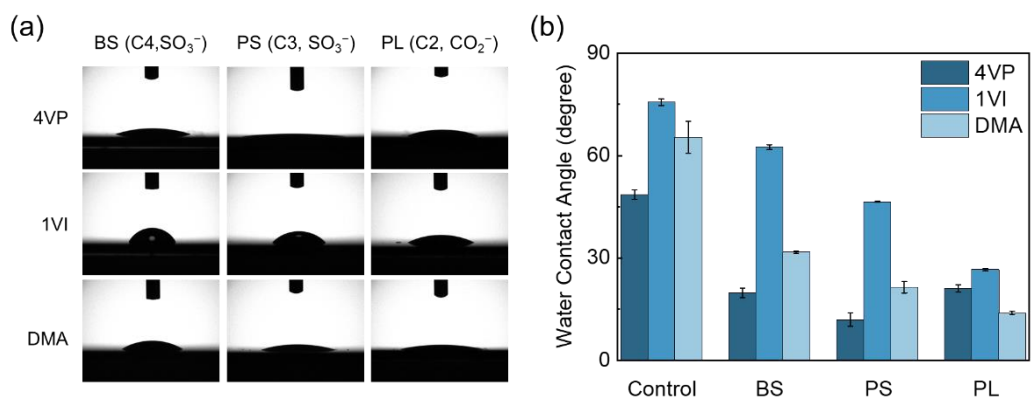

**Figure 4. Water contact angle (WCA) measurements of the zwitterionic coatings with compositional gradients.** (a) Representative images of static water droplets on the surfaces of different zwitterionic coatings, illustrating variations in surface wettability. (b) Average WCA (°) for each polymer coating. N = 4, error bar = standard deviation.

To assess the antifouling performance of the zwitterionic polymers with compositional gradients, we incubated two bacteria— the gram-negative *Pseudomonas aeruginosa* strain PAO1 and the gram-positive *Bacillus subtilis* DS2569—on coated 96-well plates for 24 hours. The quantities of the biofilms were subsequently measured using the crystal violet assay (**Figure 5**). The 1VI-PL coating demonstrated the best antifouling efficacy against PAO1, reducing the biofilm formation to 6% that on 4VP-PS and 22% that on 1VI-PS, a marked improvement especially given that 4VP-PS and 1VI-PS have been reported to have outstanding antifouling performance(*12, 15*) (**Figure 5 (a)**). The 1VI-PL was similarly fouling resistant against DS2569, leading to biofilm formation that was merely 2% that on 4VP-PS and 3% that on 1VI-PS (**Figure 5 (b)**). Notably, the 1VI-based zwitterionic polymers consistently outperformed others, regardless of the anion choice, suggesting that imidazolium-based zwitterionic chemistry is particularly effective at inhibiting biofilm formation, pointing to the potential of 1VI-PL as the next-generation anti-biofilm chemistry. Interestingly, the DS2569 biofilm growth was more pronounced on PS-derivatized zwitterionic polymers irrespective of the cation.

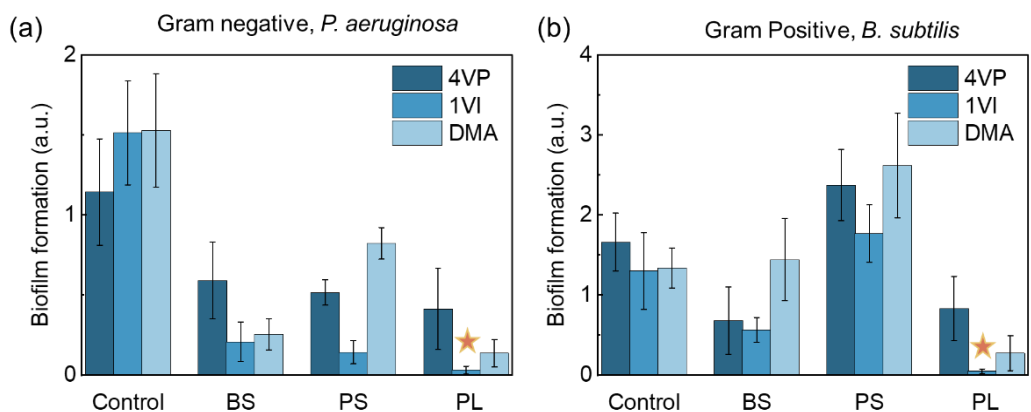

**Figure 5.** Antifouling performance of the zwitterionic polymer coatings with compositional gradients against biofilm formed by (a) gram-negative *Pseudomonas aeruginosa* (PAO1) and (b) gram-positive *Bacillus subtilis* (DS2569). N = 4, error bar = standard deviation.

**Conclusion**

In this study, we developed a library of 9 zwitterionic chemistries to illustrate molecular design rules for antifouling zwitterionic polymers. The antifouling efficacy of these coatings was assessed using gram-negative *P. aeruginosa* PAO1 and gram-positive *B. subtilis* DS2569. Among the tested polymers, 1VI-PL exhibited superior resistance to biofilm formation, reducing biofilm accumulation to as low as 2% that on previously optimized zwitterionic polymer coatings. The findings suggest that imidazolium-based zwitterionic polymers, particularly when combined with carboxyl or sulfonate groups, exhibit enhanced hydration and fouling resistance, making them ideal candidates for antifouling applications. The synthesis approach presented here is all-dry and substrate independent, thus paving the way for future applications in biomedical, marine, and industrial applications.

**Acknowledgments**

The project is sponsored by the Department of the Navy, Office of Naval Research under ONR award N00014-23-1-2189.

**Conflict of Interest**

On behalf of all authors, the corresponding author states that there is no conflict of interest.

Control Measures, and Innovative Treatment. *Microorganisms*. 2023 (10.3390/microorganisms11061614).

**Supplementary Figures**

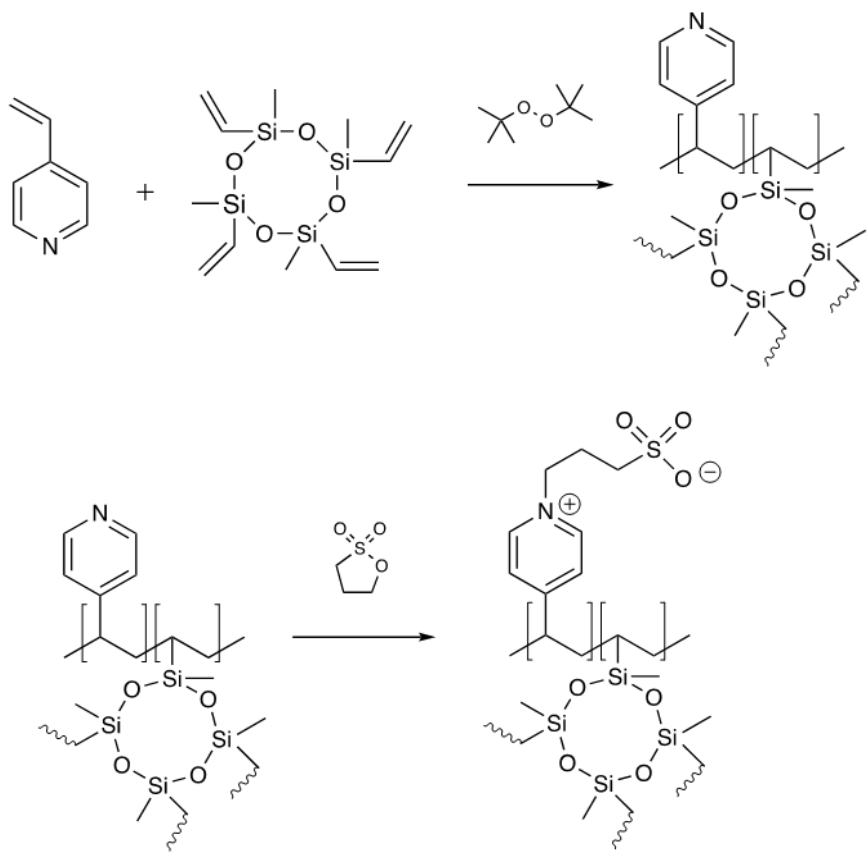

Figure S1. Representative reactions in the 2-step synthesis pathway.

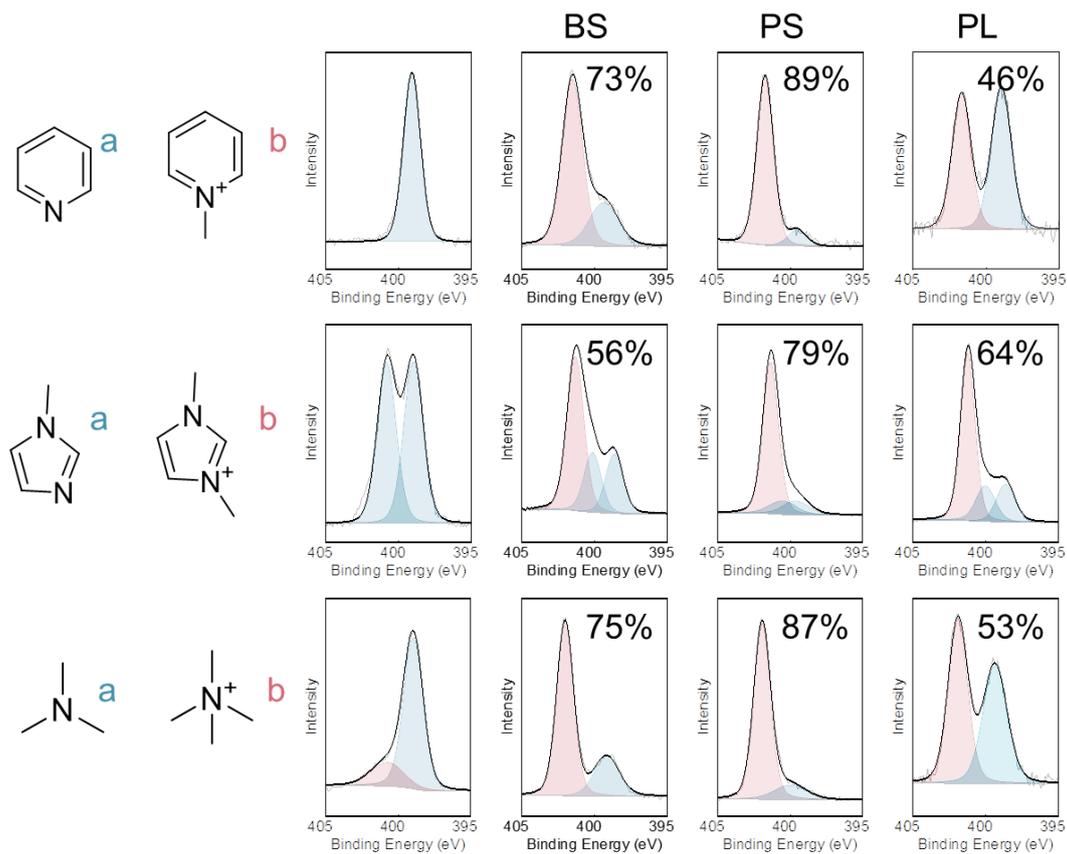

Figure S2. XPS N1s high-resolution scan of the top layer of the gradient zwitterionic polymer coatings.

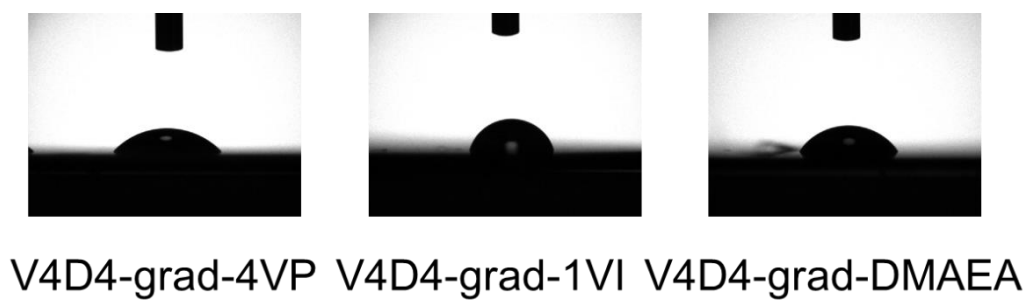

Figure S3. Optical images of a 2-μL water droplet on the gradient polymer coatings.